\documentclass[hyper]{JHEP} 

\usepackage{epsfig}

\newcommand\fverb{\setbox\pippobox=\hbox\bgroup\verb}

\newcommand\fverbdo{\egroup\medskip\noindent%

			\fbox{\unhbox\pippobox}\ }

\newcommand\fverbit{\egroup\item[\fbox{\unhbox\pippobox}]}

\newcommand\mL{\mathcal{L}}
\newcommand\mT{\mathcal{T}}
\newcommand\otheta{\overline{\theta}}
\newcommand\mF{\mathcal{F}}
\newcommand\he{\hat{e}}
\newcommand{\bB}{\mathbf{B}}
\newcommand{\bA}{\mathbf{A}}
\newcommand{\bC}{\mathbf{C}}
\newcommand{\bD}{\mathbf{D}}
\newcommand{\bG}{\mathbf{G}}
\newcommand{\bBi}{\left(\bB^{-1}
\right)}

\newcommand{\tX}{\tilde{X}}
\newcommand{\tg}{\tilde{g}}
\newcommand{\tx}{\tilde{x}}
\newcommand{\ttheta}{\tilde{\theta}}
\newcommand{\tu}{\tilde{u}}
\newcommand{\tmF}{\tilde{\mF}}
\newcommand{\tthe}{\tilde{\he}}
\newbox\pippobox

\title{Non-Relativistic Non-BPS Dp-brane}

\author{J. Kluso\v{n}
 \footnote{On leave from Masaryk University, Brno}\\
Dipartimento di Fisica \& Sezione I.N.F.N.\\
Universit\`a di Roma
``Tor Vergata'' \\
Via della Ricerca  Scientifica, 1 00133  Roma   ITALY\\
E-mail:
\email{Josef.Kluson@roma2.infn.it}}
\preprint{ \\
\hepth{0610073}}

\abstract{We construct 
non-relativistic non-BPS Dp-brane action.
Then we will study the properties of the
tachyon kink solution on its world-volume. 
We will argue that this tachyon kink
describes non-relativistic D(p-1)-brane.}


\def\bA{\mathbf{A}}

\begin{document}

\section{Introduction and Summary}
Non-relativistic string theory 
was determined some years ago in the
context of the study of the string
theories in the background with the
electric field
\cite{Seiberg:2000ms,
Gopakumar:2000na,Seiberg:2000gc,Gopakumar:2000ep,
Bergshoeff:2000ai,Harmark:2000ff}.
Then it  was soon recognised in 
\cite{Gomis:2000bd,Danielsson:2000gi,Danielsson:2000mu}
that non-relativistic string theory
is consistent sector of the string theory
with the manifest world-sheet conformal
field theory description that has
appropriate Galilean symmetry. 
For the case of strings, this can
be accomplished if we consider wound strings
in the presence of a background $B$-field
and tuning the $B$-field so that the energy coming
from the $B$-field cancels the tension of the string.
Then it was shown in \cite{Gomis:2005bj}
that the similar procedure  can be performed in 
case of Dp-branes   
\footnote{For related works, see
\cite{Brugues:2004an,Gomis:2004pw,Gomis:2005pg,
Sakaguchi:2006pg,Kamimura:2005rz}}.
Since Dp-branes are charged with respect
to Ramond-Ramond form $C_{p+1}$ 
it is possible to find the limit where
the tension of the wound Dp-brane is
cancelled by the coupling to the 
$C_{p+1}$ field. As a result we 
  obtain a world-volume kappa symmetric
action of a non-relativistic Dp-brane
\cite{Gomis:2005bj}.

We can also ask the question whether 
some symmetries of the relativistic string
theories remain  symmetries of the non-relativistic
string theories as well. 
It was shown in \cite{Kamimura:2005rz}
that in the non-relativistic string theories
one can find T-duality relation between
effective actions for non-relativistic
Dp-branes and also
one can argue that non-relativistic
D1 and D2-brane actions arise in
the dimensional reduction from 
non-relativistic M2-brane action.

On the other hand it is  well known that all 
Dp-branes can be interpreted  as
the result  of the tachyon condensation
on the world-volume of unstable D-branes
or on the world-volume of
D-brane anti-D-brane pair 
\cite{Witten:1998cd,Horava:1998jy}
\footnote{For review, see 
\cite{Sen:1999mg,Sen:2004nf}.}.
One can then ask the question whether
non-relativistic BPS Dp-branes 
can be considered as the result of
the tachyon condensation on the world-volume
of non-relativistic non-BPS Dp-brane.
The goal of this paper is to
study this problem.  

Our approach closely follows  
\cite{Gomis:2005bj} with some mild
differences. First of all we would like
to find non-relativistic non-BPS Dp-brane
action where the tachyon contribution
has the schematic form $V(T)F(
\eta^{ij}\partial_iT\partial_jT)$ with 
$F(u)\sim u^{1/2}$ for large $u$. The
reason for this requirement is that
for such a form of the  non-BPS Dp-brane 
action one
can find the tachyon kink solution 
that represents codimenison one D(p-1)-brane
\cite{Sen:2003tm}. This condition however
 implies that in the scaling limit that
 defines non-relativistic Dp-brane from 
 the relativistic one 
 the tachyon has to scale in the
 same way as the  
 world-volume scalar modes that define
the embedding of Dp-brane along
the directions spanned by the
 background Ramond-Ramond $C_{p}$
form. On the other hand if we 
try to apply procedure given 
in   \cite{Gomis:2005bj} where
these modes scale with the factor
that goes to infinity  we obtain
 that the argument  of the tachyon
potential goes to infinity that
however implies that the tachyon
potential goes to zero.
As a result  the non-BPS
Dp-brane disappears in
this definition of the non-relativistic
limit. To resolve this paradox
we use the fact that  the
non-relativistic limit can
be interpreted as the 
limit where the
world-volume scalar modes that
parametrise directions 
 transverse to the Ramond-Ramond background
are small. More precisely,
since  the
correct definition of the non-relativistic
limit of BPS Dp-brane needs background Ramond-Ramond
form $C_{p+1}$ it is natural to
split the world-volume scalar
modes to two sets, one that corresponds
to the  modes that parametrise directions
along the Ramond-Ramond form $C_{p+1}$ and
the second one that parametrises
directions transverse to it. In
the case of the ordinary BPS Dp-brane
its world-volume can span  the directions
specified by the background Ramond-Ramond $C_{p+1}$
form. However in case of a non-BPS Dp-brane
the situation is different since
there does not exist
  such a $C_{p+1}$ form. 
On the other hand 
it is well known that the non-BPS Dp-brane
couples to the Ramond-Ramond 
forms through the Wess-Zumino (WZ) term
that contains gradient of the
tachyon and the tachyon potential
\cite{Okuyama:2003wm,
Kraus:2000nj,Kennedy:1999nn,Billo:1999tv,
Takayanagi:2000rz}. 
We will argue  that thanks to the
existence of this WZ term
it is still possible to define
the correct non-relativistic non-BPS
Dp-brane as well. 

As the further support of our 
result 
we will study the tachyon kink on
the world-volume of a non-relativistic
 non-BPS Dp-brane.
Using the approach given in 
\cite{Sen:2003tm} we will construct
the  singular tachyon kink
on the world-volume of the non-relativistic
non-BPS Dp-brane 
that can be interpreted as a
lower dimensional non-relativistic 
D(p-1)-brane. This result then suggests
that we  can define non-relativistic
non-BPS Dp-brane that has the same
properties as its relativistic 
version. 

This paper is organised as follows.
In the next section (\ref{second})
we introduce the bosonic
part of the  non-BPS Dp-brane action
in type IIA or type IIB theories.  Then
in section (\ref{third}) we will consider
its non-relativistic limit. In section
(\ref{fourth}) we will study the
tachyon kink on its world-volume.
In section (\ref{fifth}) we review
the basic properties of the 
supersymmetric form of the non-BPS
Dp-brane in type IIA theory.
Finally in section (\ref{sixth}) we  
consider its  non-relativistic limit. 
\section{Review of non-BPS Dp-brane}\label{second}
In this section we review the basic facts
considering 
the bosonic part of the non-BPS Dp-brane
action  
\cite{Sen:1999md,Bergshoeff:2000dq,
Garousi:2000tr,Kluson:2000iy}
\footnote{General properties of the
tachyon effective actions were also
discussed in 
\cite{Sen:2002qa,Fotopoulos:2003yt,Sen:2003bc,
Kutasov:2003er,Niarchos:2004rw}.}. The bosonic
part of a non-BPS Dp-brane action takes the
form 
\footnote{We work in units $(2\pi\alpha')=1$.
Note also that $p$ is odd in type IIA theory while
$p$ is even in type IIB theory.}
\begin{eqnarray}\label{actb}
S&=&S_{DBI}+S_{WZ} \  , \nonumber \\
S_{DBI}&=&-\tau_p\int d^{p+1}\xi e^{-\Phi} V(T)
\sqrt{-\det\bA} \ , \nonumber \\ 
S_{WZ}&=\tau_p &\int_\Sigma  
V(T)dT\wedge 
 C \wedge e^{
F+B} \ , 
\nonumber \\
\end{eqnarray}
where 
\begin{eqnarray}
\bA_{ij}&=&\partial_i X^M
\partial_j X^N G_{MN}+
\partial_i X^M\partial_j X^NB_{MN}+
F_{ij}+
\partial_i
T\partial_j T \  ,
 \quad 
\nonumber \\ 
F_{ij}&=&\partial_i A_j
-\partial_j A_i \ ,
\nonumber \\
\end{eqnarray}
where $A_i \ ,
i,j=0,\dots,p$ and $ X^{M,N} \ ,
M,N=0,\dots,9$ are gauge and the
transverse scalar
fields on the world-volume of
the non-BPS Dp-brane and
$T$ is the
tachyon field.
$V(T)$ is the tachyon potential that
is symmetric under $T\rightarrow -T$,
has maximum at $T=0$
equal to $V(0)=1$ 
and has its minimum at $T=\pm \infty$
where it vanishes. 
$\tau_p$ is a tension of the non-BPS Dp-brane
that is related to the tension of the BPS Dp-brane
$T_p$ as $\tau_{p}=\sqrt{2}T_p$.
Finally $G_{MN},
B_{MN}$ and $\Phi$ are background
metric, $B$-field and dilaton respectively.
We restrict to the  
background with flat
Minkowski metric $G_{MN}=\eta_{MN}=
\mathrm{diag}(-1,1,\dots,1)$,
with vanishing $B$-field and 
constant dilaton $\Phi_0$. In what
follows we also include $e^{-\Phi_0}$
into the definition of $\tau_p$.  

Note also that $S_{WZ}$ given 
in (\ref{actb}) is WZ
term for a non-BPS Dp-brane that 
expresses the coupling of the
non-BPS Dp-brane to the Ramond-Ramond
forms.  In (\ref{actb})
 $\Sigma$ denotes the  world-volume
of the non-BPS Dp-brane and $C$ collects all
RR n-form gauge potentials (pulled back to the
world-volume) as
\begin{equation}
C=\oplus_n C_{(n)} \ .
\end{equation}
The form
of the WZ term was determined
 from the requirement that
the Ramond-Ramond charge of the tachyon
kink is equal to the charge of
D(p-1)-brane 
\cite{Okuyama:2003wm,
Kraus:2000nj,Kennedy:1999nn,Billo:1999tv,
Takayanagi:2000rz}. In fact it was argued in 
\cite{Okuyama:2003wm,Sen:2003tm}
that the consistency requires:
\begin{equation}
\tau_p\int_{-\infty}^{\infty}
V(T)dT=T_{p-1} \ , 
\end{equation}
where $T_{p-1}$ is a tension of 
a BPS D(p-1)-brane. 
\section{Non-relativistic
limit of the non-BPS Dp-brane}\label{third}
In this section we  define 
 non-relativistic limit of a  non-BPS Dp-brane.
As was argued in   \cite{Gomis:2005bj} the
correct definition of the  non-relativistic limit 
of BPS Dp-brane is based on the
existence of the
  constant background  Ramond-Ramond form $C_{p+1}$
  and the BPS 
 Dp-brane is extended in the  directions determined
by the orientation of 
$C_{p+1}$ in order to cancel Dp-brane tension by
the coupling to the $C_{p+1}$ field.  For non-BPS
Dp-brane there is not suitable $C_{p+1}$ form with the
rank equal to the dimension of the world-volume
of a non-BPS Dp-brane. Moreover
the form of the WZ term given
in (\ref{actb}) contains tachyon  so that
we cannot proceed exactly in the same
way as in \cite{Gomis:2005bj}. 
As we argued in the introduction we
define the non-relativistic limit as
the limit where 
the world-volume scalar modes that parametrise
directions  \emph{transverse to the 
directions spanned by
the background Ramond-Ramond form} $C_{p}$
are small. Then we also demand that the
tachyon and tachyon potential does not scale
in the non-relativistic limit. 
More precisely let us presume that the
Ramond-Ramond  field $C_{p}$ 
takes following form
\begin{equation}\label{Cbag}
C_{\mu_0\dots \mu_{p-1}}=
-(-1)^p\epsilon_{\mu_0\dots \mu_{p-1}} \ , 
\quad 
\mu,\nu=0,\dots,p-1 \ .
\end{equation}
Then (\ref{Cbag})
suggests following scaling 
of the world-volume variables
\begin{eqnarray}\label{nrelaB}
X^\mu &=& x^\mu \ , \nonumber \\ 
X^a&=&  \lambda X^a \ , \nonumber \\
\tau_p&=&\lambda^{-2}
\tau_{NRp} \ , \nonumber \\
F_{ij}&=&
\lambda f_{ij}  \ , \nonumber \\
A_i&=& \lambda W_i \ , 
\nonumber \\
T&=&
T \ ,
\nonumber \\
\end{eqnarray}
where $a,b=p,\dots,9$. 
Note that we have also presumed that
the world-sheet gauge field $A_i$ scales
in the same way as the transverse modes.
The non-relativistic limit is defined
as limit $\lambda\rightarrow 0$. 

Now for (\ref{nrelaB}) 
 the matrix $\bA$ takes the form
\begin{eqnarray}\label{bBlambda}
\bA_{ij}&=&
g_{ij}+\lambda f_{ij}+
\partial_i T\partial_j T+\lambda^2 x^ax^b g_{ab}
\equiv 
\bB_{ij}+\lambda \bC_{ij}+\lambda^2 \bD_{ij} \ , 
\nonumber \\
\bB_{ij}&=& g_{ij}+\partial_iT\partial_j T \ , 
g_{ij}=\partial_i x^\mu\partial_j x^\nu\eta_{\mu\nu} \ , 
\nonumber \\
 \bC_{ij}&=& f_{ij} \ , 
\bD_{ij}=X^aX^b g_{ab} \  . 
\nonumber \\ 
\end{eqnarray}
Inserting (\ref{bBlambda})
into the DBI part of the tachyon effective
action and considering the terms up to orders
$\lambda^2$ we 
obtain the non-relativistic
Dp-brane in the form 
\begin{eqnarray}\label{NRac}
S_{NR}&=&-\frac{\tau_{NRp}}{\lambda^2}
\int d^{p+1}\xi V(T)\sqrt{-\det \bB}
-\nonumber \\
&-&
\frac{\tau_{NRp}}{2}\int d^{p+1}\xi
V(T)\sqrt{-\det\bB}\left[\bBi^{ij}
\partial_j X^a\partial_i X^b g_{ab}-
\right.\nonumber \\
 &-& \left.
\frac{1}{2}
\bBi^{ik}f_{kj}\bBi^{jl}f_{li}+O(\lambda^2)
\right] \ . \nonumber \\
\end{eqnarray}
Before we turn to the  non-relativistic
limit in the WZ term let us study in more
details the matrix $\bB$. Its form
suggests that it is natural to temporarily introduce
the notation
\begin{equation}
Y^I\equiv (x^\mu,T) \ , \quad
Y^p=T \ , I,J=0,\dots,p \ 
\end{equation}
so that we can rewrite $\bB$ into the suggestive
form
\begin{equation}
\bB_{ij}=\partial_i Y^I\partial_j Y^J g_{IJ}
\equiv E^I_j E^J_j\eta_{IJ} \ , 
\eta_{pp}=1 \ .  
\end{equation}
Using this notation it is easy to see
that the  divergent contribution in 
(\ref{NRac})  can be written as
\begin{equation}\label{Actdiv}
S_{div}=\frac{\tau_{NRp}}{\lambda^2}
\int \frac{1}{(p+1)!}V(T)
E^{I_0}\wedge \dots \wedge E^{I_p} \ , 
\end{equation}
where $E^I\equiv E^I_j d\xi^j$.

Let us study the non-relativistic
limit in the WZ term. We  insert the 
pull-back of Ramond-Ramond background
form (\ref{Cbag}) into it and we
obtain 
\begin{eqnarray}\label{SWZdiv}
S_{WZ}
&=&-\frac{\tau_{NRp}}{\lambda^2}
\int  V(T)dT \frac{1}{p!}(-1)^p
\epsilon_{\mu_0\dots \mu_{p-1}}
dX^{\mu_0}\wedge
\dots \wedge dX^{\mu_{p-1}}=
\nonumber \\
&-&\frac{\tau_{NRp}}{\lambda^2}
\int V(T) \frac{1}{p!}
\epsilon_{\mu_0\dots \mu_{p-1}}
dX^{\mu_0}\wedge
\dots \wedge dX^{\mu_{p-1}}\wedge dT=
\nonumber \\
&-&\frac{\tau_{NRp}}{\lambda^2(p+1)!}
\int V(T)\epsilon_{I_0\dots I_p}
E^{I_0}\wedge \dots \wedge E^{I_p} \ .
\nonumber \\
\end{eqnarray}
If we add (\ref{SWZdiv})  to
(\ref{Actdiv}) we obtain 
that  all divergent contributions
in the full non-relativistic non-BPS Dp-brane
cancel. As the result we obtain 
 finite non-relativistic    non-BPS Dp-brane action 
\begin{eqnarray}\label{SNRpfin}
S_{NRp}^{fin}&=&-\frac{\tau_{NRp}}{2}\int d^{p+1}\xi
V(T)\sqrt{-\det\bB}\left(\bBi^{ij}
\partial_j X^a\partial_i X^b g_{ab}-
\right.\nonumber \\
 &-& \left.
\frac{1}{2}
\bBi^{ik}f_{kj}\bBi^{jl}f_{li}
\right) \ . \nonumber \\
\end{eqnarray}
In the next section we will study the
tachyon kink on the world-volume 
of the action (\ref{SNRpfin}).
\section{Tachyon kink on
the world-volume of non-relativistic
non-BPS Dp-brane}\label{fourth}
In this section we will study 
the tachyon kink solution on the
world-volume of the   
non-relativistic  non-BPS Dp-brane
following  
\cite{Sen:2003tm}
\footnote{For generalisation of this
approach to the curved background
see \cite{Kluson:2005fj,Kluson:2005hd}}.

As in \cite{Sen:2003tm} we start
with the following ansatz
\begin{eqnarray}\label{ansT}
T&=&f(a(\xi^p- t(\xi^\alpha))) \ , 
\nonumber \\
W_\alpha&=&w_\alpha(\xi^\alpha) \ , 
\quad 
W_p=0 \ , 
\nonumber \\ 
x^\mu&=& x^\mu(\xi^\alpha)  \ , \quad
\mu=0,\dots,p-1 \ , 
\nonumber \\
X^a&=& x^a(\xi^\alpha) \ , 
 \quad a=p,\dots,9 \ , 
\nonumber \\
\end{eqnarray}
where $\xi^\alpha,\alpha,\beta=0,1,\dots, p-1$ 
are coordinates tangential to the
 the world-volume of the kink.
The function
$f$ introduced in 
(\ref{ansT}) satisfies
\begin{equation}
f(-u)=-f(u) \ , \quad f'(u)>0 \ \quad 
\forall u \ , \quad f(\pm)=\pm \infty
\end{equation}
but is otherwise an arbitrary function of
the argument $u$. $a$ is a constant
that we shall take to $\infty$ in the end. 
In this limit we have $T=\infty$ for
$\xi^p>0$ and $T=-\infty$ for $\xi^p<0$. 

The  first goal of our analysis 
is tho shown that the action (\ref{SNRpfin})
evaluated on the ansatz (\ref{ansT})
reproduces in the limit
$a\rightarrow \infty$  the 
 action for non-relativistic
D(p-1)-brane 
\begin{eqnarray}\label{abnrp-1}
S_{NR(p-1)}&=&-
\frac{T_{NR(p-1)}}{2}\int d^{p}\xi
\sqrt{-\det \tg }(\tg^{\alpha\beta}
\partial_\alpha x^a\partial_\beta
 x^b g_{ab}-
\frac{1}{2}
\tg^{\alpha\beta}f_{\beta\gamma}
\tg^{\gamma\delta}f_{\delta\alpha}
) \ , \nonumber \\
\end{eqnarray}
where  
\begin{eqnarray}
\tg&=&\partial_\alpha x^\mu
\partial_\beta x^\nu\eta_{\mu\nu} \ , \quad
\mu \ , \nu=0,\dots,p-1 \ , 
\nonumber \\
f_{\alpha\beta}&=&\partial_\alpha w_\beta-
\partial_\beta w_\alpha \ , 
\quad \alpha,\beta=0,\dots,p-1 \ 
\end{eqnarray}
and $x^a \ , a,b=p,\dots,9$ parametrise
positions of non-relativistic Dp-brane
transverse to its world-volume. 
For letter purposes we also determine
the equations of motion that arise
from (\ref{abnrp-1})
\begin{eqnarray}\label{eqxmuB}
-\frac{1}{2}\partial_\alpha\left[
\eta_{\mu\nu}
\partial_\beta x^\nu \tg^{\beta\alpha}
\sqrt{-\det\tg}
\left(\tg^{\gamma\delta}
\partial_\delta x^a\partial_\gamma x^b g_{ab}-
\frac{1}{2}
\tg^{\gamma\delta}f_{\delta
\omega}\tg^{\omega\sigma}f_{\sigma\gamma}\right)\right]-
\nonumber \\
-\partial_\alpha[\sqrt{-\det\tg}\tg^{\alpha\beta}
\eta_{\mu\nu}\partial_\gamma x^\nu\tg^{\gamma \delta}
\partial_\delta x^a\partial_\beta x^bg_{ab}]
+\nonumber \\
+\partial_\alpha [
\sqrt{-\det\tg}
\tg^{\alpha\beta}f_{\beta\sigma}\tg^{\sigma\omega}
f_{\omega\delta}\tg^{\delta\gamma}
\partial_\gamma x^\nu
\eta_{\nu\mu}]=0 \  \nonumber \\
\end{eqnarray}
that is the equation of motion for $x^\mu$.
On the other hand the 
 equation of motion for $x^a$
takes much simpler  form  
\begin{equation}\label{eqXB}
\partial_\alpha [
\sqrt{-\det \tg}\tg^{\alpha \beta}\partial_\beta x^b
g_{ba}]=0 \ .
\end{equation}
Finally, we determine the equation
of motion for $w_\alpha$
\begin{equation}\label{eqWiB}
\partial_\beta[\sqrt{-\det \tg}
\tg^{\beta \delta}f_{\delta \gamma}\tg^{\gamma \alpha}]=0
\ . 
\end{equation}
Let us now  insert the 
ansatz  (\ref{ansT}) into the
matrix  $\bB$ defined in
(\ref{bBlambda})
\begin{equation}\label{bB}
\bB_{ij}=
\left(\begin{array}{cc}
\tg_{\alpha\beta}+a^2f'^2\partial_\alpha t\partial_\beta t & 
-a^2f'^2\partial_\alpha t \\
-a^2 f'^2\partial_\beta t & a^2f'^2 \\
\end{array}\right) \  
\end{equation}
and evaluate its determinant
\begin{equation}\label{detB}
\det\bB=
\left|\begin{array}{cc}
\bB_{\alpha\beta}-\bB_{\alpha y}
\bB_{pp}^{-1}\bB_{p\beta} & 0 \\
\bB_{p\beta} & \bB_{pp}\\
\end{array}\right|=a^2 f'^2\det \tg_{\alpha\beta} \   \ 
\end{equation}
and its inverse matrix $\bB^{-1}$
\begin{eqnarray}\label{Bin}
\bBi^{pp}&=&\tg^{\alpha\beta}\partial_\alpha
t\partial_\beta t \ , \quad
\bBi^{p\alpha}=\partial_\beta t \tg^{\beta\alpha} \ , 
\nonumber \\
\bBi^{\alpha p}&=&\tg^{\alpha\beta}\partial_\beta t \ , 
\quad 
\bBi^{\alpha\beta}=\tg^{\alpha\beta} \  .
\nonumber\\
\end{eqnarray}
Note that the form of the matrix $\bB^{-1}$ is
exact for all $a$. 
If we now insert 
(\ref{ansT}), (\ref{detB})
and (\ref{Bin}) into 
(\ref{SNRpfin}) and consider
the limit of large $a$ we
obtain
\begin{eqnarray}\label{SNRpfinev}
S_{NRp}^{fin}&=&\frac{T_{NR(p-1)}}{2}\int d^p\xi
\sqrt{-\det\tg}\left[\tg^{\alpha\beta}
\partial_\alpha x^a\partial_\beta x^b g_{ab}-
\right.\nonumber \\
 &-& \left.
\frac{1}{2}
\tg^{\alpha\beta}
f_{\beta\gamma}\tg^{\gamma\delta}
f_{\delta\alpha}\right]
 \ , \nonumber \\
\end{eqnarray}
where we have used the fact that
\begin{equation}\label{Ttbps}
\frac{\tau_{NRp}}{2}
\int d\xi^p V(f(a\xi^p))af'(\xi^p)=
\frac{\tau_{NRp}}{2}
\int dmV(m)=\frac{T_{NR(p-1)}}{2} \ .  
\end{equation}
Comparing (\ref{abnrp-1})
with (\ref{SNRpfinev}) we obtain
the result that the tachyon kink
reproduces the non-relativistic
D(p-1)-brane action. We also see that $t(\xi)$
 does not appear in 
 (\ref{SNRpfinev})
 and hence its interpretation is unclear
 at present. 

Note however that this result
does not prove that the dynamics of the
kink is governed by the action
(\ref{abnrp-1}). To do this we 
have to show, following
\cite{Sen:2003tm}, that any solution of
the equations of motion derived from
(\ref{abnrp-1}) will produce a solution
of the equations of motions derived
from (\ref{SNRpfin}) under the identification
(\ref{ansT}).

In order to establish this correspondence
we firstly determine the equations of motions
that follow from (\ref{SNRpfin}). By varying
the action (\ref{SNRpfin}) with respect to
$T$ we obtain 
\begin{eqnarray}\label{Teq}
& &\frac{1}{2}
V'(T)\sqrt{-\det\bB}
\left(\bBi^{ij}
\partial_j X^a\partial_i X^b g_{ab}-
\frac{1}{2}
\bBi^{ik}f_{kj}\bBi^{jl}f_{li}\right)-
\nonumber \\
&-&\frac{1}{2}\partial_i
\left[\partial_j T\bBi^{ji}
\sqrt{-\det\bB}[
\bBi^{ij}
\partial_j X^a\partial_i X^b g_{ab}-
\frac{1}{2}
\bBi^{ik}f_{kj}\bBi^{jl}f_{li}]\right]-
\nonumber \\
&-&\partial_l [V(T)\sqrt{-\det\bB}\bBi^{il}
\partial_k T \bBi^{kj}\partial_i X^a\partial_j X^b]
+\nonumber \\
&+&\partial_m[V(T)
\sqrt{-\det\bB}
\bBi^{im}\partial_n T
\bBi^{nk}f_{kj}\bBi^{jl}f_{li}]=0
\ . 
\nonumber \\
\end{eqnarray}
In the same way we determine the equations
of motion for $x^\mu$
\begin{eqnarray}\label{eqxmu}
&-&\frac{1}{2}\partial_i\left[
V(T)\eta_{\mu\nu}
\partial_j x^\nu \bBi^{ji}
\sqrt{-\det\bB}
\left(\bBi^{ij}
\partial_j X^a\partial_i X^b g_{ab}-
\right.\right. \nonumber \\
&-&\left.\left.
\frac{1}{2}
\bBi^{ik}f_{kj}\bBi^{jl}f_{li}\right)\right]-
\nonumber \\
&-&\partial_m[V(T)\bBi^{im}
\eta_{\mu\nu}\partial_kx^\nu\bBi^{kj}
\partial_j X^a\partial_i X^bg_{ab}]
+\nonumber \\
&+&\partial_m[V(T)
\sqrt{-\det\bB}
\bBi^{im}\eta_{\mu\nu}
\partial_n x^\nu
\bBi^{nk}f_{kj}\bBi^{jl}f_{li}]=0 \  
\nonumber \\
\end{eqnarray}
and the  equations of motion
 for $X^a$ 
\begin{equation}\label{eqXa}
\partial_i [V(T)
\sqrt{-\det\bB}\bBi^{ij}\partial_jX^b
g_{ba}]=0 \ .
\end{equation}
Finally, we determine the equation
of motion for $W_i$
\begin{equation}\label{eqWi}
\partial_j[
V(T)\sqrt{-\det\bA}
\bBi^{jk}f_{kl}\bBi^{li}]=0
\ . 
\end{equation}
Now we will 
solve the equations of motions
(\ref{Teq}),(\ref{eqXa}),(\ref{eqWi})
with the ansatz (\ref{ansT}).
 We start our discussion 
 with the equation
(\ref{eqXa}). Inserting
(\ref{detB})\ , (\ref{Bin})
and (\ref{ansT}) to (\ref{eqXa})  we
obtain
\begin{eqnarray}\label{eqXaf}
& &\partial_p[V(T)\sqrt{-\det\bB}
\bBi^{x\alpha}\partial_\alpha X^b
g_{ba}]+\partial_\beta[V(T)
\sqrt{-\det\bB}\bBi^{\beta\alpha}
\partial_\alpha X^b g_{ba}]=
\nonumber \\
&=&a \partial_p[Vf']\sqrt{-\det\tg}
\partial_\beta t \tg^{\beta\alpha}
\partial_\alpha x^bg_{ba}
+a\partial_\beta[V f']\sqrt{-\det\tg}\tg^{\beta\alpha}
\partial_\alpha x^b g_{ba}+\nonumber \\
&+& af'V(T)\partial_\beta[
\sqrt{-\det\tg}\tg^{\beta\alpha}
\partial_\alpha x^b g_{ba}]=
af'V(T)\partial_\beta[
\sqrt{-\det\tg}\tg^{\beta\alpha}
\partial_\alpha x^b g_{ba}]
\ , \nonumber \\
\end{eqnarray} 
where we have used the fact that
\begin{equation}
\partial_\alpha[V(f)af']=
-\partial_\alpha t\partial_p[V(f)
af'] \ . 
\end{equation}
Now the final form of the equation
(\ref{eqXaf}) implies that the equation
of motion (\ref{eqXa}) is obeyed for
the ansatz (\ref{ansT}) on condition
that $x^a$ obeys (\ref{eqXB}).
In the same way we proceed with 
(\ref{eqWi}). For $i=p$ we obtain
\begin{eqnarray}\label{ip}
& &\partial_j[
V(T)\sqrt{-\det\bB}
\bBi^{jk}f_{kl}\bBi^{lp}]=
- \partial_p[Vaf']\sqrt{-\det\tg}
\partial_\beta t\tg^{\beta\alpha}
f_{\alpha\delta}\tg^{\delta\gamma}
\partial_\gamma t
\nonumber \\
&-& af'V(T)
\partial_\alpha[\sqrt{-\det\tg}\tg^{\alpha\gamma}
f_{\gamma\beta}\tg^{\beta\delta}\partial_\delta t]=
-af'V(T)\sqrt{-\det\tg}\tg^{\alpha\gamma}
f_{\gamma\beta}\tg^{\beta\delta}\partial_\alpha
\partial_\delta t -
\nonumber \\
&-& af'V(T)
\partial_\alpha[\sqrt{-\det\tg}\tg^{\alpha\gamma}
f_{\gamma\beta}\tg^{\beta\delta}]\partial_\delta t=
-af'V(T)
\partial_\alpha[\sqrt{-\det\tg}\tg^{\alpha\gamma}
f_{\gamma\beta}\tg^{\beta\delta}]\partial_\delta t \ , 
\nonumber \\
\end{eqnarray}
where in the first step we have
used the fact that the expression
proportional to $\partial_p[Vaf']$ vanishes
thanks to fact that  $\tg^{\alpha\beta}$ is symmetric
while  $f_{\alpha\beta}$
is antisymmetric.  
In the second step we  have again used  the 
antisymmetry of $f$ so that
\begin{eqnarray}
\tg^{\alpha\gamma}
f_{\gamma\beta}\tg^{\beta\delta}\partial_\alpha
\partial_\delta t=0 \ .  \nonumber \\ 
\end{eqnarray}
Then the final form of the
equation (\ref{ip})
implies that  the ansatz
(\ref{ansT}) solves the
 equation
of motion (\ref{eqWi}) 
on condition that 
the modes $w_\alpha,x^a,x^\mu$
obey (\ref{eqWiB}). 

 We can proceed
in the same way 
in case when $i$
in (\ref{eqWi})
 is equal to $\beta$ and we get 
\begin{eqnarray}
& &\partial_j[
V(T)\sqrt{-\det\bB}
\bBi^{jk}f_{kl}\bBi^{l\beta}]=
 af'V 
\partial_\alpha[
\sqrt{-\det\tg}\tg^{\alpha\gamma}
f_{\gamma\delta}\tg^{\delta\beta}] \ . 
\nonumber \\
\end{eqnarray}
This result again implies that
the ansatz (\ref{ansT}) obeys
the equation of motion (\ref{eqWi})
on condition that $w_\alpha,x^\mu,x^a$ solve
 (\ref{eqWiB}).

Finally we will analyse the
 tachyon equation of motion  (\ref{Teq}).
To simplify the calculation we 
use the fact that
\begin{equation}
\bBi^{ij}\bB_{j p}=\delta^i_p \ , \quad 
\bB_{p j}\bBi^{ji}=\delta_p^i
\end{equation}
and then using (\ref{bB}) we obtain
\begin{eqnarray}\label{exBi}
\bBi^{ip}-\bBi^{i\alpha}\partial_\alpha t=\frac{1}{a^2f'^2}
\delta^i_p \ .
\nonumber \\
\end{eqnarray}
Let us start with  the
expression on the first and
the second line in 
(\ref{Teq}) 
\begin{eqnarray}\label{Teqfe}
& &\frac{1}{2}V'(T)\sqrt{-\det\bB}
\left(\bBi^{ij}
\partial_j X^a\partial_i X^b g_{ab}-
\frac{1}{2}
\bBi^{ik}f_{kj}\bBi^{jl}f_{li}\right)-
\nonumber \\
&-&\frac{1}{2}\partial_i
\left[\partial_j T\bBi^{ji}
\sqrt{-\det\bB}\left[
\bBi^{ij}
\partial_j X^a\partial_i X^b g_{ab}-
\frac{1}{2}
\bBi^{ik}f_{kj}\bBi^{jl}f_{li}\right]\right]=
\nonumber \\
&=&\frac{1}{2}V'(T)\sqrt{-\det\bB}
\left(\bBi^{ij}
\partial_j X^a\partial_i X^b g_{ab}-
\frac{1}{2}
\bBi^{ik}f_{kj}\bBi^{jl}f_{li}\right)-
\nonumber \\
&-&\frac{1}{2}\partial_i\left[V(f)\frac{1}{a^2f'^2}
\delta^i_p
\sqrt{-\det\bB}\left[
\bBi^{ij}
\partial_j X^a\partial_i X^j g_{ij}-
\frac{1}{2}
\bBi^{ik}f_{kj}\bBi^{jl}f_{li}\right]\right]=0 \ , 
\nonumber \\
\end{eqnarray}
where we have used
(\ref{exBi}). Moreover, using
(\ref{exBi}) in remaining
terms in (\ref{Teq}) we get
that all these terms vanish.
We see that  the ansatz
 (\ref{ansT}) solves
the equation of motion (\ref{Teq})
for arbitrary $t(\xi)$. 
 In fact this is satisfactory result since
 we work with the action where 
 the world-volume diffeomorphism invariance
is not fixed and  hence $\xi^p$ is 
 equivalent to $\xi^p+t(\xi)$. This
fact also implies that it is natural
to consider 
$t(\xi)$ as a  parameter of 
the gauge diffeomorphism transformations
and hence it has no  physical significance.  
 
Let us outline the results derived
above. We have shown that the dynamics of the
 massless modes given in (\ref{ansT}) is governed
 by the action (\ref{abnrp-1}). This result
 supports the interpretation of the tachyon
 kink as the  D(p-1)-brane. 
\section{Supersymmetric 
Relativistic  Non-BPS Dp-brane}
\label{fifth}
In this section we define supersymmetric
form of a non-relativistic non-BPS Dp-brane.
We firstly review the  basic facts
 about non-BPS D-branes in Type IIA theory,
 following \cite{Sen:1999md}
\footnote{For non-BPS D-brane in Type IIB theory
 the situation is basically the
 same with difference in
chirality of the Majorana-Weyl fermions.}.
Let $ \xi^{i} , i,j=0,...,p $ are 
world-volume coordinates on a D-brane.
On a non-BPS Dp-brane world-volume we
have a $32$ component anti-commuting
field $\theta$ that transforms
as a Majorana spinor of the 10 dimensional
Lorentz group. We also denote 
$\Gamma^M$ the ten dimensional
gamma-matrices that 
 can be chosen real
by taking charge conjugation matrix $C=\Gamma_0$.
These matrices obey the anti-commutation
relation
\begin{equation}
\{\Gamma^M,\Gamma^N\}=2\eta^{MN}
\end{equation} with 
$\eta^{MN}=(-1,1,...,1)$.
 We also introduce $\Gamma_{11}=
\Gamma_0...\Gamma_9 , (\Gamma_{11})^2=1$.
 
The supersymmetric generalisation of the
the DBI part of the  effective action
 for a  non-BPS Dp-brane  has a form
\cite{Sen:1999md,Bergshoeff:2000dq,Garousi:2000tr,Kluson:2000iy}: 
\begin{equation}\label{action}
 S_{DBI}=-\tau_p\int d^{p+1}\xi V(T)\sqrt{-\det
(\mathcal{G}_{ij}+\mathcal{F} 
_{ij}+\partial_i T
\partial_j T )}=-\int \mL_{DBI} \ , 
\end{equation}
where 
\begin{equation}\label{formPi}
\Pi^M_{i}=\partial_{i}X^M+i\overline{\theta}
\Gamma^M\partial_{i}\theta \ , \quad  \mathcal{G}_{ij}=
\eta_{MN}\Pi^M_{i}\Pi^N_{j} 
\end{equation}
and
\begin{equation}
\mathcal{F}_{ij}=F_{ij}-
b_{ij} \ , \quad F_{ij}=\partial_i A_j-
\partial_j A_i \ . 
\end{equation}
There also exists the supersymmetric
generalisation of the WZ term
\begin{equation}\label{WZsup}
S_{WZ}=\tau_p\int_\Sigma  V(T)dT  
  \wedge \Omega_{p}=\int_\Sigma \mL_{WZ} \ .  
 \end{equation}
The form $\Omega_p$ expresses the coupling
of the non-BPS Dp-brane to the Ramond-Ramond fields
 \cite{Bergshoeff:2000dq}. In particular,
its bosonic part denotes the pull back of the
Ramond-Ramond fields and the combination
of the field strength $F$. This part vanishes
for vanishing Ramond-Ramond background but 
there is a part of $\Omega_p$ that survives
even in the absence of any Ramond-Ramond
background
\cite{Cederwall:1996pv,Aganagic:1996pe,
Cederwall:1996ri,Bergshoeff:1996tu}.
  Since 
 the explicit form of
$\Omega_p$ is complicated we
introduce, following 
\cite{Gomis:2005bj} $(p+1)$
form $h_{p+1}$  such that
\begin{equation}
h_{p+1}=d\Omega_{p} \ . 
\end{equation}
For type IIA non-BPS Dp-branes
(p-odd) the forms $b,h_{p+1}$
are equal to
\begin{eqnarray}
b&=&-i\otheta\Gamma_{11}\Gamma_Md\theta
\wedge   
(\Pi^M-\frac{i}{2}\otheta \Gamma^Md\theta) \ , 
\nonumber \\
h_{p+1}&=&-(-1)^n i d\overline{\theta}\wedge 
\mT_{p-1}\wedge d\theta  \ , n=\frac{p-1}{2} \ ,
\nonumber \\ 
\end{eqnarray}
where $\mT_{p-1}$ is $p-1$ form. To define
it, we introduce the formal sum of differential
forms
\begin{equation}
\mT_A=\sum_{p'=even}
\mT_{p'}=
e^{\mF}C_A \ ,
\end{equation}
where 
\begin{equation}
C_A=\Gamma_{11}+
\frac{1}{2!}\psi^2+\frac{1}{4!}
\Gamma_{11}\psi^4+\frac{1}{6!}
\psi^6+\dots 
\end{equation}
and
\begin{equation}
\psi=\Pi^M\Gamma_M \ . 
\end{equation}
One can shown that the non-BPS
Dp-brane action is invariant
under rigid supersymmetry variations
but it is  not invariant under local
$\kappa$ symmetry transformations
\cite{Sen:1999md}. The absence of this
local symmetry implies that we
cannot gauge away one half of the
fermions  and hence
the world-volume theory contains the
correct number of massless degrees
of freedom \cite{Sen:1999md}.

We can generalise the analysis presented
in the previous section and switch
on one more coupling in the world-volume
of non-BPS Dp-brane consistent with all
symmetries of the action
\cite{Gomis:2005bj}.
 From the
space-time point of view, this is equivalent
to turning on a closed  Ramond-Ramond
p-form $C_p$ which does not modify the
supergravity equations of motion. Then
the total non-BPS Dp-brane action takes
the form
\begin{equation}
S=-\int d^{p+1}\xi
\mL_{DBI}+\int \mL_{WZ}
+\int \mL_{C_p} \ , 
\end{equation}
where
\begin{equation}
\mL_{C_p}=\tau_p V(T)dT\wedge f^*C_p \ , 
\end{equation}
where $f^*$ 
 is the pullback
of $C_p$ on the world-volume
of non-BPS Dp-brane. 
\section{Non-relativistic non-BPS 
Dp-brane with presence of fermions}\label{sixth}
We derive the action
for non-relativistic non-BPS Dp-brane 
from the supersymmetric form of the relativistic
non-BPS Dp-brane action.
We again closely follow \cite{Gomis:2005bj}.
 
The non-relativistic limit
 is obtained by decoupling
some charged light degrees
of freedom that obey non-relativistic
dispersion relation from the full
relativistic theory. This decoupling
is achieved in the similar
way as in  section (\ref{second}).
In other words we demand that the
scalar modes that parametrise directions
transverse to the background 
Ramond-Ramond p-form $C_p$ 
are small. Considering
fermionic degrees of freedom
we will scale them in the way
that is equivalent to the
scaling introduced in
\cite{Gomis:2004pw,Gomis:2005bj}
\begin{eqnarray}\label{nrelaS}
X^\mu &=& x^\mu \ , \nonumber \\ 
X^a&=&   \lambda X^a \ , \nonumber \\
\tau_p&=&\lambda ^{-2}
\tau_{NRp} \ , \nonumber \\
F_{ij}&=&
\lambda  f_{ij}  \ , \nonumber \\
A_i&=& \lambda W_i \ , 
\nonumber \\
T&=&T \ ,  
\nonumber \\
\theta&=&\theta_-+
\lambda\theta_+ \ ,
\nonumber \\
C_{\mu_0\dots \mu_{p-1}}&=&
-\epsilon_{\mu_0\dots \mu_{p-1}} \ ,
\nonumber  \\
\end{eqnarray}
where $X^M$ has been split in
$X^\mu \ , \mu=0,\dots,p-1$
 and $X^a,a=p,\dots,9$. The $X^\mu$ are
coordinates of target space parallel
to the orientation of
$C_p$  and $X^a$ are transverse to it.
Finally $\lambda$ is scaling parameter
that is sent to zero in the end. 

  The scaling of fermions
depends on the splitting of the
fermions under the matrix $\Gamma_*$:
\begin{equation}
\Gamma_*\theta_\pm=\pm\theta_\pm \ .
\end{equation}
The matrix $\Gamma_*$ is defined
as
\begin{equation}
\Gamma_*=(-1)^{n+1}\Gamma_{0\dots p-1}
\Gamma^{n+1}_{11} \  , 2n=p-1 \  
\end{equation}
and it  obeys following
 relations
\begin{eqnarray}
\otheta_\pm&=&\pm(-1)^{\frac{p+1}{2}}
\otheta_\pm \Gamma_* \ , \quad
\Gamma_*\Gamma_{11}=-\Gamma_{11}
\Gamma_* \ , \nonumber \\
\Gamma_*\Gamma^{\mu}&=&(-1)^{\frac{p+1}{2}}
\Gamma^\mu \Gamma_* \ , 
\quad 
\Gamma_*\Gamma^a=(-1)^{\frac{p-1}{2}}
\Gamma^a\Gamma_* \  .
\nonumber \\
\end{eqnarray}
Note also that $\Gamma_*^2=\mathrm{I}$. 
Now we insert (\ref{nrelaS})
to the  supertranslation
1-form given in (\ref{formPi}) and
we obtain
\begin{eqnarray}\label{Pformscal}
\Pi^\mu &=&
\he^\mu+i\lambda^2\otheta_+\Gamma^\mu
d\theta_+ 
 \ , \quad
\he^\mu=e^\mu+i\otheta_-\Gamma^\mu d\theta_- \ , 
\quad e^\mu=dx^\mu \ ,
\nonumber \\ 
\Pi^a&=& 
\lambda  u^a \ ,  \quad 
u^a= dx^a+2i\otheta_+\Gamma^a d\theta_- \ ,
\quad 
x^a=X^a+i\otheta_-\Gamma^a\theta_+ \ .
\nonumber \\
\end{eqnarray}
In the same way we determine the scaling 
of the
form $\mF$ 
\begin{eqnarray}\label{mFscal}
\mF&=&
  \lambda \mF^{(1)}+
  \lambda^3\mF^{(3)}  \ , 
\nonumber \\
\mF^{(1)}&=&
f+\frac{i}{2}
(\otheta_+\Gamma_{\mu}\Gamma_{11}
d\theta_-+i\otheta_-\Gamma_\mu\Gamma_{11}
d\theta_-)(\he^\mu-\frac{i}{2}\otheta_-
\Gamma^\mu d\theta_-)+
\nonumber \\
&+&i
\otheta_-\Gamma_{a}\Gamma_{11}
d\theta_-
(u^a-\frac{i}{2}\otheta_-\Gamma^a
d\theta_+-\frac{i}{2}\otheta_+
\Gamma^a d\theta_+) \ , \nonumber \\
\mF^{(3)}&=&
\frac{1}{2}
(\otheta_+\Gamma_{\mu}\Gamma_{11}
d\theta_-+\otheta_-\Gamma_\mu\Gamma_{11}
d\theta_-)
\otheta_+\Gamma^\mu d\theta_+
\nonumber \\
&+&i\otheta_+\Gamma_{11}\Gamma_a
d\theta_+
(u^a-\frac{i}{2}\otheta_-\Gamma^a
d\theta_+-\frac{i}{2}\otheta_+
\Gamma^a d\theta_+) \ .  \nonumber \\
\end{eqnarray}
Following \cite{Gomis:2005bj} 
we will keep $\lambda$ 
small but finite  in the intermediate computations
and only send $\lambda $ to zero in the
end. For that reason we keep explicitly
terms in the action that scale as 
 $\frac{1}{\lambda}$ (that are divergent)
and terms 
that are independent
on $\lambda$ (that are finite). We also drop
terms that scale as  $\lambda$ since they cannot contribute
when we take the limit $\lambda\rightarrow 0$
at the end of the analysis. 

More precisely, if we insert 
(\ref{Pformscal}) into
$\mathcal{G}$ we obtain
\begin{eqnarray}\label{mGs}
\mathcal{G}_{ij}&=&
 \Pi^M_i\Pi^N_j\eta_{MN}=
 \bG^{(0)}_{ij}
+\lambda^2\bG^{(2)}_{ij}+
O(\lambda^3) \ , \nonumber \\
\bG_{ij}^{(0)}&=&\he_i^\mu \he_j^\nu\eta_{\mu\nu} \ , 
\quad
 \bG_{ij}^{(2)}=
2i\he^\mu \otheta_+\Gamma^\nu
d\theta_+\eta_{\mu\nu}+u_i^au_j^b\delta_{ab}
\ , \nonumber \\
\end{eqnarray}
where we have restricted to
the terms up to order $\lambda^2$. 
Then if we insert 
(\ref{mFscal}) and (\ref{mGs})
  into the DBI part of
the tachyon effective action 
(\ref{action}) we obtain
supersymmetric form of the non-relativistic
non-BPS Dp-brane action  in Type IIA theory
\begin{eqnarray}\label{Siia}
S_{NR,DBI}&=&-\frac{\tau_{NRp}}{\lambda^2}\int d^{p+1}
\xi V(T)\sqrt{-\det\bB}-
\nonumber \\
&-&\frac{\tau_{NRp}}{2}\int d^{p+1}
\xi V(T)\sqrt{-\det\bB}\left(\bBi^{ij}\bG_{ji}^{(2)}-
\right.
\nonumber \\
&-& \left.
\frac{1}{2}\bBi^{ij}\mF_{jk}^{(1)}\bBi^{kl}
\mF_{li}^{(1)}\right)+O(\lambda) \  ,  \nonumber \\
\end{eqnarray}
where
\begin{equation}
\bB_{ij}=\bG_{ij}^{(0)}+
\partial_iT\partial_j T=
\he_i^\mu \he_j^\nu\eta_{\mu\nu}+
\partial_iT\partial_j T
 \ . 
\end{equation}
In order to carefully treat with the
divergent term in (\ref{Siia})
we perform the
same simplification  as in section (\ref{second}) 
and  temporarily write
\begin{eqnarray}\label{Bijs}
\bB_{ij}&=&
\he_i^I \he_j^J\eta_{IJ} \ , \quad
I,J=0,\dots,p   \ , \nonumber \\
\he^{I}_i&=& \he^{\mu}_i \ , \quad  \mathrm{for} \  
\mu=I=0,\dots,p-1 \ , \quad
\he^p_i=\partial_iT \ . 
\nonumber \\
\end{eqnarray}
With the help of (\ref{Bijs})
 we can  rewrite the divergent
term in (\ref{Siia}) as
\begin{eqnarray}
&-&\frac{\tau_{NRp}}{\lambda^2}
\int d^{p+1}\xi V(T)
\sqrt{-\det\bB}= \nonumber \\
&=&
\frac{\tau_{NRp}}{\lambda^2 (p+1)!}
\int V(T) \epsilon_{I_0\dots I_p}
\he^{I_0}\wedge \dots \wedge \he^{I_p}
\equiv \frac{1}{\lambda^2}
\int d^{p+1}\xi \mL_{DBI}^{div} 
 \ . \nonumber \\
\end{eqnarray}
Now  we insert (\ref{nrelaS})
into the WZ term. As in 
\cite{Gomis:2005bj} we easily obtain
\begin{equation}
\tau_{p}h_{p+1}=
\frac{\tau_{NRp}}{\lambda^2}
h_{p+1}^{(2)}+
\tau_{NRp}h_{p+1}^{(0)}+
O(\lambda^2) \ . 
\end{equation}
The superficially divergent term
can be determined as in \cite{Gomis:2005bj}
and takes the form
\begin{eqnarray}
h_{p+1}^{(2)}&=&
-id\otheta_-\wedge \frac{1}{(p-1)!}
\he^{\mu_1}\wedge \dots \wedge
\he^{\mu_{p-1}}\epsilon_{\mu_1\dots
\mu_{p-1}\nu}\Gamma^\nu \wedge d\theta_-=
\nonumber \\
&=& -\epsilon_{\nu \mu_1\dots 
\mu_{p-1}}\frac{1}{(p-1)!}
d\he^\nu \wedge \he^{\mu_1}
\wedge \dots \wedge e^{\mu_{p-1}} \ . 
\nonumber \\
\end{eqnarray}
Now we  consider following expression
\begin{eqnarray}\label{canc}
d(d^{p+1}\xi \mL^{div})&=&
-d(d^{p+1}\xi \mL^{div}_{DBI})+
d( V(T)dT\wedge \Omega_{p})^{div}=
\nonumber \\
&=&\tau_{NRp}
V(T)\left[d\left(\frac{1}{ (p+1)!}
 \epsilon_{I_0\dots I_p}
\he^{I_0}\wedge \dots \wedge \he^{I_p}\right)
+dT\wedge h_{p+1}\right]=
\nonumber \\
&=&\tau_{NRp}
V(T)\left[\frac{1}{p!}
 \epsilon_{J I_1 \dots I_p}
d\he^{J}\wedge \he^{I_1}\wedge \dots \wedge \he^{I_p}
+dT\wedge h_{p+1}\right]=0
\nonumber \\
\end{eqnarray}
using the fact that
\begin{equation}
dV\wedge \he^p=
\frac{dV}{dT}\he^p\wedge \he^p=0 \ , 
\quad
d(dT)=0  \ 
\end{equation}
and also
\begin{eqnarray}
&-&dT\wedge \epsilon_{\nu \mu_1\dots
\mu_{p-1}}\frac{1}{(p-1)!}
d\he^{\nu}\wedge \he^{\mu_1}\wedge\dots
\wedge \he^{\mu_{p-1}}= \nonumber \\
&-&(-1)^{(p+1)}
 \epsilon_{\nu \mu_1\dots
\mu_{p-1}}\frac{1}{(p-1)!}
d\he^{\nu}\wedge \he^{\mu_1}\wedge\dots
\wedge \he^{\mu_{p-1}}\wedge dT=\nonumber \\
&=&-\frac{1}{p!}
\epsilon_{J I_1\dots I_p}
d\he^{J}\wedge \he^{I_1}\wedge\dots
\wedge \he^{I_p} \ , \nonumber \\
\end{eqnarray}
where we have used  the  definition of the
exterior derivative $d$ and also
the fact that for type IIA non-BPS
Dp-brane $p$ is odd.  

As the last term in
(\ref{canc})
  involves only
the terms with fermions
\cite{Gomis:2005bj}, this
cancellation removes the terms with fermions
in $\mL_{DBI}^{div}$. There remains 
the pure bosonic term in $\mL^{div}_{DBI}$
that is
\begin{equation}
-d^{p+1}\xi\mL_{DBI,bos}^{div}=
\tau_{NRp}V(T)\frac{1}{(p+1)!}
\epsilon_{I_0\dots I_p}e^{I_0}\wedge \dots
\wedge e^{I_p} \ .
\end{equation}
This term can be cancelled by turning on
a closed Ramond-Ramond form $C_p$ that
gives an contribution to the WZ term
\begin{eqnarray}
\int \mL_{C_p}^{div}&=&
-\tau_{NRp}\int V(T)dT\wedge \frac{1}{p!} \epsilon_{\mu_0\dots
\mu_{p-1}}
dx^{\mu_0}\wedge\dots\wedge  dx^{\mu_{p-1}}=
\nonumber \\
&-& \tau_{NRp}\int V(T)\frac{1}{(p+1)!}
\epsilon_{I_0\dots I_p}
e^{I_0}\wedge \dots \wedge e^{I_p}
\ . \nonumber \\
\end{eqnarray} 
Then  we   obtain the finite part of
the supersymmetric generalisation of the
non-relativistic non-BPS Dp-brane action
in the form
\begin{eqnarray}\label{actnpbssusy}
S_{NR}&=&
-\tau_{NRp}
\int d^{p+1}\xi
V(T)\sqrt{-\det\bB}
\left[i \bBi^{ij}\otheta_+\he^\mu_i
\Gamma^\nu\eta_{\mu\nu} \partial_j\theta_+
+\right. \nonumber \\
&+&\left. 
\frac{1}{2}\bBi^{ij}u_i^a u_j^bg_{ab}-
\frac{1}{4}\bBi^{ij}\mF_{jk}^{(1)}\bBi^{kl}
\mF_{li}^{(1)}\right] +\nonumber \\
&+& \tau_{NRp}
\int V(T)dT\wedge \Omega_{p}^{(0)} \ ,
\nonumber \\
\end{eqnarray}
where $\Omega^{(0)}_p$ is the non-relativistic
WZ term that has the same form 
as in case of non-relativistic BPS D(p-1)-brane. 

Our goal is to shown that the world-volume action
for non-relativistic D(p-1)-brane in Type IIA theory
anises from the action (\ref{actnpbssusy}) as
the tachyon kink. Recall that 
the action for non-relativistic D(p-1)-brane
in type IIA theory has the form
\cite{Gomis:2005bj}
\begin{eqnarray}\label{actbssusy}
S_{NR}^{BPS}&=&
-T_{NR(p-1)}
\int d^p\xi
\sqrt{-\det \tg}
\left(i \tg^{\alpha\beta}\otheta_+\he^\mu_\alpha
\Gamma^\nu\eta_{\mu\nu} \partial_\beta\ttheta_+
+\frac{1}{2}\tg^{\alpha\beta}\tu_\alpha^a \tu_\beta^b
g_{ab}-\right.\nonumber \\
\nonumber \\
&-&\left.\frac{1}{4}\tg^{\alpha\beta}\tmF_{\beta\gamma}^{(1)}
\tg^{\gamma\delta}
\tmF_{\delta \alpha}^{(1)}\right) +T_{NR(p-1)}
\int  \Omega_{p}^{(0)} \ ,
\nonumber \\
\end{eqnarray}
where 
\begin{eqnarray}
\tg_{\alpha\beta}&=&
\tthe^{\mu}_\alpha \tthe_{\beta}^\nu \eta_{\mu\nu} \ ,
\quad 
\tthe_\beta^\mu=
\tilde{e}^\mu+i\overline{\ttheta}_-\Gamma^\mu 
d\ttheta_- \ , 
\quad \tilde{e}^\mu=d\tx^\mu \ ,
\nonumber \\
\tu^a &=& d\tx^a+2i\overline{\ttheta}_+\Gamma^a d\ttheta_-  
, \quad 
\tx^a=\tX^a+i\overline{\ttheta}_-\Gamma^a\ttheta_+ \ ,
\nonumber \\
\end{eqnarray}
and where $\mF^{(1)}$ has the same form as
in (\ref{mFscal}) with the gauge field $\tilde{w}_\alpha$.

In order to show that the tachyon kink on the
world-volume of the supersymmetric non-relativistic
non-BPS Dp-brane describes the 
non-relativistic BPS D(p-1)-brane we should
perform the same analysis as 
in section (\ref{second}). 
As the first step we consider 
the ansatz
\begin{eqnarray}\label{ansTsusy}
T&=&f(a(\xi^p-t(\xi^\alpha))) \ , \nonumber \\
W_\alpha(\xi^p,\xi^\alpha)&=&
\tilde{w}_\alpha(\xi^\alpha)  \ , \quad 
\quad W_p=0 \ , \nonumber \\
 x^\mu(\xi^p,\xi^\alpha)&=&
\tx^\mu(\xi^\alpha) \ , \quad
X^a(\xi^p,\xi^\alpha)=\tx^a(\xi^\alpha) \ , 
\nonumber \\
\theta_+(\xi^p,\xi^\alpha)&=&
\ttheta_+(\xi^\alpha) \ , \quad 
\theta_-(\xi^p,\xi^\alpha)=\ttheta_-(\xi^\alpha) \ 
\nonumber \\
\end{eqnarray}
and insert it to the 
matrix  $\bB$ so that we obtain 
\begin{equation}\label{bBsusy}
\bB_{ij}=
\left(\begin{array}{cc}
\tg_{\alpha\beta}+a^2f'^2\partial_\alpha t
\partial_\beta t  & -a^2f'^2\partial_\alpha t\\
-a^2f'^2\partial_\beta t & a^2 f'^2 \\
\end{array}\right) \ . 
\end{equation}
Then we also get   
\begin{equation}\label{detBsusy}
\det\bB=a^2 f'^2\det \tg_{\alpha\beta} \   \ 
\end{equation}
and
\begin{eqnarray}\label{Binsusy}
\bBi^{pp}&=&\tg^{\alpha\beta}\partial_\alpha
t\partial_\beta t \ , \quad
\bBi^{p\alpha}=\partial_\beta t \tg^{\beta\alpha} \ , 
\nonumber \\
\bBi^{\alpha p}&=&\tg^{\alpha\beta}\partial_\beta t \ , 
\quad 
\bBi^{\alpha\beta}=\tg^{\alpha\beta} \ .   
\nonumber\\
\end{eqnarray}
Then it is easy to see that
when we insert 
 (\ref{ansTsusy}),(\ref{bBsusy}),(\ref{detBsusy})
and (\ref{Binsusy}) into (\ref{actnpbssusy}),
 perform the integration over $\xi^p$ 
and use the relation (\ref{Ttbps})
we obtain the non-relativistic
D(p-1)-brane action  (\ref{actbssusy}).

To show the complete equivalence we should
check that the ansatz (\ref{ansTsusy}) solves
the equations of motion that arise
from (\ref{actnpbssusy}). However from the
form of the ansatz (\ref{ansTsusy}) and
corresponding matrix $\bB$ it is clear
that the ansatz  solves these equations
on conditions that the massless modes
given there solve the equations of motion
that follow from (\ref{actbssusy}).

\section*{Acknowledgements}

This work
 was supported in part by the Czech Ministry of
Education under Contract No. MSM
0021622409, by INFN, by the MIUR-COFIN
contract 2003-023852, by the EU
contracts MRTN-CT-2004-503369 and
MRTN-CT-2004-512194, by the INTAS
contract 03-516346 and by the NATO
grant PST.CLG.978785.



\newpage

\begin{thebibliography}{40}


\bibitem{Seiberg:2000ms}
  N.~Seiberg, L.~Susskind and N.~Toumbas,
 \emph{``Strings in background 
electric field, space/time noncommutativity  and a
new noncritical string theory,''}
  JHEP {\bf 0006}, 021 (2000)
  [arXiv:hep-th/0005040].

\bibitem{Gopakumar:2000na}
  R.~Gopakumar, J.~M.~Maldacena, S.~Minwalla and A.~Strominger,
\emph{``S-duality and noncommutative gauge theory,''}
  JHEP {\bf 0006}, 036 (2000)
  [arXiv:hep-th/0005048].

\bibitem{Seiberg:2000gc}
  N.~Seiberg, L.~Susskind and N.~Toumbas,
\emph{``Space/time non-commutativity and causality,''}
  JHEP {\bf 0006}, 044 (2000)
  [arXiv:hep-th/0005015].

\bibitem{Gopakumar:2000ep}
  R.~Gopakumar, S.~Minwalla, N.~Seiberg and A.~Strominger,
 \emph{``OM theory in diverse dimensions,''}
  JHEP {\bf 0008}, 008 (2000)
  [arXiv:hep-th/0006062].

\bibitem{Bergshoeff:2000ai}
  E.~Bergshoeff, D.~S.~Berman, J.~P.~van der Schaar and P.~Sundell,
 \emph{``Critical fields on 
the M5-brane and noncommutative open strings,''}
  Phys.\ Lett.\ B {\bf 492}, 193 (2000)
  [arXiv:hep-th/0006112].

\bibitem{Harmark:2000ff}
  T.~Harmark,
  \emph{``Open branes in space-time 
non-commutative little string theory,''}
  Nucl.\ Phys.\ B {\bf 593}, 76 (2001)
  [arXiv:hep-th/0007147].
  


\bibitem{Gomis:2000bd}
  J.~Gomis and H.~Ooguri,
 \emph{``Non-relativistic closed string theory,''}
  J.\ Math.\ Phys.\  {\bf 42} (2001) 3127
  [arXiv:hep-th/0009181].

\bibitem{Danielsson:2000gi}
  U.~H.~Danielsson, A.~Guijosa and M.~Kruczenski,
 \emph{``IIA/B, wound and wrapped,''}
  JHEP {\bf 0010} (2000) 020
  [arXiv:hep-th/0009182].

\bibitem{Danielsson:2000mu}
  U.~H.~Danielsson, A.~Guijosa and M.~Kruczenski,
\emph{``Newtonian gravitons and 
D-brane collective coordinates in wound string
theory,''}
  JHEP {\bf 0103}, 041 (2001)
  [arXiv:hep-th/0012183].





\bibitem{Brugues:2004an}
  J.~Brugues, T.~Curtright, J.~Gomis and L.~Mezincescu,
 \emph{``Non-relativistic 
strings and branes as non-linear realizations of  Galilei
groups,''}
  Phys.\ Lett.\ B {\bf 594} (2004) 227
  [arXiv:hep-th/0404175].


\bibitem{Gomis:2004pw}
  J.~Gomis, K.~Kamimura and P.~K.~Townsend,
 \emph{``Non-relativistic superbranes,''}
  JHEP {\bf 0411} (2004) 051
  [arXiv:hep-th/0409219].


\bibitem{Gomis:2005pg}
  J.~Gomis, J.~Gomis and K.~Kamimura,
\emph{``Non-relativistic 
superstrings: A new soluble sector of AdS(5) x S**5,''}
  JHEP {\bf 0512} (2005) 024
  [arXiv:hep-th/0507036].

\bibitem{Gomis:2005bj}
  J.~Gomis, F.~Passerini, T.~Ramirez and A.~Van Proeyen,
\emph{``Non relativistic Dp branes,''}
  JHEP {\bf 0510} (2005) 007
  [arXiv:hep-th/0507135].

\bibitem{Sakaguchi:2006pg}
  M.~Sakaguchi and K.~Yoshida,
\emph{``Non-relativistic 
AdS branes and Newton-Hooke superalgebra,''}
  arXiv:hep-th/0605124.


\bibitem{Kamimura:2005rz}
  K.~Kamimura and T.~Ramirez,
 \emph{``Brane dualities in 
non-relativistic limit,''}
  JHEP {\bf 0603} (2006) 058
  [arXiv:hep-th/0512146].




\bibitem{Witten:1998cd}
  E.~Witten,
  \emph{``D-branes and K-theory,''}
  JHEP {\bf 9812} (1998) 019
  [arXiv:hep-th/9810188].


\bibitem{Horava:1998jy}
  P.~Horava,
 \emph{``Type IIA D-branes, K-theory, and matrix theory,''}
  Adv.\ Theor.\ Math.\ Phys.\  {\bf 2}, 1373 (1999)
  [arXiv:hep-th/9812135].




\bibitem{Sen:1999mg}
  A.~Sen,
\emph{``Non-BPS states and 
branes in string theory,''}
  arXiv:hep-th/9904207.

\bibitem{Sen:2004nf}
  A.~Sen,
 \emph{``Tachyon dynamics in open string theory,''}
  Int.\ J.\ Mod.\ Phys.\ A {\bf 20} (2005) 5513
  [arXiv:hep-th/0410103].








\bibitem{Sen:1999md}  
A.~Sen,
  \emph{``Supersymmetric world-volume action 
for non-BPS D-branes,''}
  JHEP {\bf 9910} (1999) 008
  [arXiv:hep-th/9909062].

\bibitem{Bergshoeff:2000dq}
  E.~A.~Bergshoeff, M.~de Roo, 
T.~C.~de Wit, E.~Eyras and S.~Panda,
\emph{``T-duality and actions for non-BPS D-branes,''}
  JHEP {\bf 0005} (2000) 009
  [arXiv:hep-th/0003221].

\bibitem{Garousi:2000tr}
  M.~R.~Garousi,
 \emph{ ``Tachyon couplings on 
non-BPS D-branes and Dirac-Born-Infeld action,''}
  Nucl.\ Phys.\ B {\bf 584} (2000) 284
  [arXiv:hep-th/0003122].

\bibitem{Kluson:2000iy}
  J.~Kluson,
\emph{ ``Proposal for non-BPS D-brane action,''}
  Phys.\ Rev.\ D {\bf 62} (2000) 126003
  [arXiv:hep-th/0004106].  
HEP-TH 0004106;




\bibitem{Sen:2002qa}
  A.~Sen,
  \emph{``Time and tachyon,''}
  Int.\ J.\ Mod.\ Phys.\ A {\bf 18} 
(2003) 4869
  [arXiv:hep-th/0209122].

\bibitem{Fotopoulos:2003yt}
  A.~Fotopoulos and A.~A.~Tseytlin,
  \emph{``On open superstring 
partition function in 
inhomogeneous rolling tachyon
  background,''}
  JHEP {\bf 0312}, 025 (2003)
  [arXiv:hep-th/0310253].

\bibitem{Sen:2003bc}
  A.~Sen,
  \emph{``Open and closed 
strings from unstable D-branes,''}
  Phys.\ Rev.\ D {\bf 68}, 106003 (2003)
  [arXiv:hep-th/0305011].

\bibitem{Kutasov:2003er}
  D.~Kutasov and V.~Niarchos,
  \emph{``Tachyon effective actions 
in open string theory,''}
  Nucl.\ Phys.\ B {\bf 666}, 56 (2003)
  [arXiv:hep-th/0304045].

\bibitem{Niarchos:2004rw}
  V.~Niarchos,
  \emph{``Notes on tachyon 
effective actions and 
Veneziano amplitudes,''}
  Phys.\ Rev.\ D {\bf 69}, 106009 (2004)
  [arXiv:hep-th/0401066].




\bibitem{Okuyama:2003wm}
  K.~Okuyama,
  \emph{``Wess-Zumino term
in tachyon effective action,''}
  JHEP {\bf 0305} (2003) 005
  [arXiv:hep-th/0304108].


\bibitem{Kraus:2000nj}
  P.~Kraus and F.~Larsen,
  \emph{``Boundary string
field theory of the DD-bar system,''}
  Phys.\ Rev.\ D {\bf 63} (2001) 106004
  [arXiv:hep-th/0012198].

\bibitem{Kennedy:1999nn}
  C.~Kennedy and A.~Wilkins,
\emph{``Ramond-Ramond couplings on
brane-antibrane systems,''}
  Phys.\ Lett.\ B {\bf 464} (1999) 206
  [arXiv:hep-th/9905195].

\bibitem{Billo:1999tv}
  M.~Billo, B.~Craps and F.~Roose,
\emph{ ``Ramond-Ramond couplings of non-BPS D-branes,''}
  JHEP {\bf 9906} (1999) 033
  [arXiv:hep-th/9905157].

\bibitem{Takayanagi:2000rz}
  T.~Takayanagi, S.~Terashima and T.~Uesugi,
  \emph{``Brane-antibrane action
from boundary string field theory,''}
  JHEP {\bf 0103} (2001) 019
  [arXiv:hep-th/0012210].

\bibitem{Sen:2003tm}
  A.~Sen,
  \emph{``Dirac-Born-Infeld 
action on the tachyon kink and vortex,''}
  Phys.\ Rev.\ D {\bf 68} (2003) 066008
  [arXiv:hep-th/0303057].



\bibitem{Kluson:2005fj}
  J.~Kluson,
\emph{``Tachyon kink on 
non-BPS Dp-brane in the general background,''}
  JHEP {\bf 0510} (2005) 076
  [arXiv:hep-th/0508239].

\bibitem{Kluson:2005hd}
  J.~Kluson,
\emph{``Note about 
tachyon kink in nontrivial background,''}
  JHEP {\bf 0508} (2005) 032
  [arXiv:hep-th/0506250].



\bibitem{Cederwall:1996pv}
  M.~Cederwall, A.~von Gussich, B.~E.~W.~Nilsson and A.~Westerberg,
\emph{``The Dirichlet super-three-brane in ten-dimensional type IIB
  supergravity,''}
  Nucl.\ Phys.\ B {\bf 490}, 163 (1997)
  [arXiv:hep-th/9610148].
  
\bibitem{Aganagic:1996pe}
  M.~Aganagic, C.~Popescu and J.~H.~Schwarz,
 \emph{``D-brane actions with local kappa symmetry,''}
  Phys.\ Lett.\ B {\bf 393}, 311 (1997)
  [arXiv:hep-th/9610249].
  
\bibitem{Cederwall:1996ri}
  M.~Cederwall, A.~von Gussich, B.~E.~W.~Nilsson, P.~Sundell and A.~Westerberg,
 \emph{``The Dirichlet super-p-branes in ten-dimensional type IIA and IIB
  supergravity,''}
  Nucl.\ Phys.\ B {\bf 490}, 179 (1997)
  [arXiv:hep-th/9611159].
  
\bibitem{Bergshoeff:1996tu}
  E.~Bergshoeff and P.~K.~Townsend,
  \emph{``Super D-branes,''}
  Nucl.\ Phys.\ B {\bf 490}, 145 (1997)
  [arXiv:hep-th/9611173].



\end{thebibliography}
\end{document}